\documentclass[twocolumn,english,aps,prb,floatfix,preprintnumbers,showpacs,amsfonts,amssymb]{revtex4}
\usepackage[T1]{fontenc}
\usepackage[latin1]{inputenc}
\usepackage{amsmath}
\usepackage{graphicx}
\usepackage{amssymb}

\makeatletter

\providecommand{\LyX}{L\kern-.1667em\lower.25em\hbox{Y}\kern-.125emX\@}

\usepackage{graphicx}
\usepackage{graphicx}
\usepackage{amscd}
\usepackage{bm}

\usepackage{babel}
\makeatother
\begin{document}

\title{NMR linewidth and Skyrmion localization in quantum Hall ferromagnets}

\author{A. Villares Ferrer, R.L. Doretto, and A.O. Caldeira }

\address{Departamento de Física da Matéria Condensada, Instituto de F\'{\i}sica
Gleb Wataghin, Universidade Estadual de Campinas, Caixa Postal 6165,
Campinas-SP 13083-970, Brazil}

\date{\today{}}

\begin{abstract}
The non-monotonic behavior of the NMR signal linewidth in the 2D quantum
Hall system is explained in terms of the interplay between skyrmions
localization, due to the influence of disorder, and the non-trivial
temperature dependent skyrmion dynamics\textbf{. }
\end{abstract}


\maketitle

\section{Introduction}

The ground state of the two-dimensional electron gas (2DEG) at filling
factor one ($\nu =1$) is a spin-polarized state \cite{girvin} in
which all electrons completely fill the lowest Landau level with spin
up polarization (quantum Hall ferromagnet). The low lying charged
excitation of the 2DEG is called skyrmion, or charged spin texture,
and carries an unusual spin distribution \cite{skyrmions-first}.
Pictorially, this distribution can be viewed as a configuration in
which the spin field points down at a given position and smoothly
rotates as one moves radially outwards from that point, until all
the spins are polarized as in the ground state. These non-trivial
objects are topologically stable, with its size -the number of reversed
spins- fixed by the competition between the Coulomb and Zeeman interactions
\cite{sky-size}. The existence of this many body electronic state
was confirmed by Barrett and coworkers \cite{skyevidence} who performed
optically pumped nuclear magnetic resonance (OPNMR) measurements of
the $^{71}$Ga nuclei located in an n-doped GaAs multiple quantum
well. As the nuclei of the quantum wells are coupled to the the spins
of the 2DEG via the isotropic Fermi contact interaction, the Knight
shift ($K_{s}$) from the NMR signal gives information about the spin
polarization of the 2DEG. It was reported that $K_{s}$ symmetrically
drops around $\nu =1$, showing that the charged excitation of the
quantum Hall ferromagnet involves a large number of spin-flips \cite{skyevidence}
in contrast with the expected behavior for non-interacting electrons.
Since then, heat capacity, magnetoabsorption and magnetotransport
measurements \cite{skyheat,skymagabs1,skytrans1} have been performed
in order to understand the main features of this novel state, however,
valuable information can be still extracted from the NMR experiments. 

Recently, new OPNMR measurements \cite{skyloc1} of the $^{71}$Ga
nuclei in n-doped GaAs/Al$_{0.1}$Ga$_{0.9}$As multiple quantum wells
were performed at temperatures ($T\approx 0.3$ K) lower than the
ones considered in Ref. \onlinecite{skyevidence}. In Ref. \onlinecite{skyloc1},
Khandelwal \textit{et al.} observed that the Knight shift data around
$\nu =1$ consist of \textbf{}a {}``tilted plateau'', in contrast
to the symmetrical fall at higher temperatures. As NMR is a local
probe, it was \textbf{}argued that the existence of the tilted plateau
is related to the localization of the skyrmions. In addition, the
full width at half maximum (FWHM) of the OPNMR signal has a non-monotonic
temperature dependence at $\nu \approx 1$, increasing when the temperature
is lowered up to a critical value and then, falling down. It is well
known \cite{slichter} that the linewidth of the NMR measurements
is affected by the dynamics of the nuclei (\textit{motional narrowing
effect}). However, as the $^{71}$Ga nuclei are spatially fixed in
the GaAs quantum wells, any observed dynamical effect must be related
to the dynamics of 2DEG. In particular, if $\nu \approx 1$, those
effects should be related to the skyrmion dynamics. In this scenario,
the behavior of the FWHM can be understood as an evolution of the
skyrmion dynamics from the \textit{motional-narrowed} to the \textit{frozen}
regimes, as the temperature decreases \cite{skyloc1}. This non-monotonic
behavior of the linewidth was previously observed by Kuzma \emph{et
al.} \cite{skyloc1ter} when the filling factor of the 2DEG was slightly
changed around $\nu =1/3$. Later on, in order to investigate the
evolution of the spectral line shape as function of $T$, Sinova and
coworkers\cite{nmrlineshape} simulated the NMR signal of $^{71}$Ga
nuclei when $\nu \approx 1$. As at low $T$ it is expected that the
short length scale correlations contain the ground state features,
the 2DEG in Ref. \onlinecite{nmrlineshape} was described by an square
skyrmion lattice\cite{skycrystal} that randomly jumps -as a whole-
over an average distance. Despite the fact that the formation of a
pinned Wigner crystal has never been observed in the range of temperatures
where the experiments\cite{skyloc1,skyloc1ter} have been performed,
the approach used in Ref. \onlinecite{nmrlineshape} reproduced the
qualitative behavior of the FWHM near the completely frozen situation.
However, as far as we know, no explicit correspondence has been established
between the temperature dependence of the skyrmion dynamics, for the
whole range of experimentally available temperatures, and the line
shape of the NMR experiments. This is precisely the main goal of this
paper.

Here, instead of trying a fully quantum mechanical approach for the
Fermi contact interaction between $^{71}$Ga nuclei and the 2DEG for
$\nu \approx 1$ we will use a semiclassical theory, in which the
skyrmion center of mass is treated as a true dynamical variable\cite{skyavf}
to which a temperature dependent diffusion constant is associated\cite{skyr-cooper}.
This way of facing the problem will allow us to study the whole range
of temperature, namely from the complete delocalized situation to
the frozen limit. 

We organize this paper as follows. In section II the model used to
describe the skyrmion dynamics and the localization by disorder are
presented and the dependence of the linewidth with the temperature
is calculated. Section III is devoted to analyze the change in the
energy of the skyrmions when impurities are taken into account. Finally
in Sec. IV we present the final results and our conclusions.

\section{The linewidth temperature dependence}

To start with we recall that the FWHM in the magnetic resonance experiment
is proportional to the induced phase shift experienced by the nuclei
every time they interact with a fluctuating field. More precisely,
\textbf{}the FWHM is equal to \textbf{$T_{2}^{-1}$}, where $T_{2}$
is the \textit{transversal relaxation time} \cite{slichter}. For
the Ga nuclei in the quantum wells, the fluctuating field is given
by the deviation from the completely polarized state provided by moving
skyrmions. Therefore, the spins of the $^{71}$Ga nuclei will precess
with an extra phase over its normal precession (in the presence of
the completely polarized state) every time a skyrmion passes near
by. This change in phase can be calculated as $\delta \phi =\pm \; \gamma _{n}\; \overline{n}\; \tau $
\cite{slichter}, where $\gamma _{n}$ is the giromagnetic ratio for
the $^{71}$Ga nuclei, $\overline{n}$ is the {}``fluctuating''
field associated to the skyrmion magnetization and $\tau $ is the
interaction time. The mean square dephasing $\langle \Delta \phi ^{2}\rangle $
after $n$ $\tau $-intervals of time will be $n\, (\gamma _{n}\, \overline{n}\, \tau )^{2}$.
Here $n=2DtN_{fs}/S$ is the number of skyrmions passing by the nucleus
in a time $t$, with $N_{fs}$ being the number of free skyrmions,
$D$ the diffusion constant and $\mathcal{S}$ the quantum well surface
area. As a result, \begin{equation}
\langle \Delta \phi ^{2}\rangle =\frac{{2DN_{fs}}}{\mathcal{S}}(\gamma _{n}\; \overline{n}\; \tau )^{2}t.\label{eq:msquare}\end{equation}

Now, if we define $T_{2}$ as the time for a group of spins in phase
at $t=0$ get about one radian out of step \cite{slichter} we find
that \begin{equation}
T_{2}^{-1}=\frac{{2DN_{fs}}}{\mathcal{S}}(\gamma _{n}\; \overline{n}\; \tau )^{2}.\label{eq:t2primeira}\end{equation}
 On the other hand, the time interval $\tau $ can be estimated from
the diffusion constant assuming that the length in which the skyrmions
effectively interact with the nucleus is about one lattice constant
$a$. In this way $\tau \sim a^{2}/2D$ and \begin{equation}
T_{2}^{-1}=\frac{{a^{4}}}{2SD}N_{fs}(\gamma _{n}\; \overline{n})^{2}.\label{eq:tesegunda}\end{equation}

Now, in order to make the temperature dependence of expression (\ref{eq:tesegunda})
explicit, we need to know how the skyrmion diffusion constant changes
with the temperature. From the semiclassical point of view the study
of the skyrmion dynamics was performed\cite{skyavf,skyr-cooper} starting
from a generalized non-linear sigma model, which has the skyrmion
as the lowest lying charged excitation\cite{skyrmions-first}. \textbf{}Within
this approach the origin of the non trivial skyrmion dynamics can
be understood in terms of its interaction with the thermal bath of
spin wave\textbf{s.} This kind of interaction can be correctly treated
using the well known collective coordinate method\cite{rajaraman}.
This technique promotes the center of mass of the skyrmion to a true
dynamical variable and, as a final result, provide us with a Hamiltonian
in which the skyrmion momentum is coupled to the generalized spin
wave momentum. The effective equation of motion for a single skyrmion
was obtained using the Feynman-Vernon functional approach\cite{feynmanvernon}
(by tracing out the magnons' degrees of freedom) and corresponds to
a Brownian particle with temperature dependent damping and diffusion
constants. \textbf{}The explicit temperature dependence of the skyrmion
diffusion coefficients\cite{skyr-cooper} is given by \begin{equation}
D=\overline{D}T^{3}\exp \left(-2g/T\right),\label{eq:difcooper}\end{equation}
where the Zeeman gap ($g=g^{*}\mu _{B}B$) will be measured in Kelvin.

In expression (\ref{eq:tesegunda}) not only the diffusion constant
but also the number of free, or moving, skyrmions are temperature
dependent quantities. In fact, as it was pointed out by Nederveen
and Nazarov\cite{nederveen}, the donors situated on a layer located
at a distance $d$ from the 2DEG (inside the potential barriers) generate
a random attractive potential (disorder) in which the skyrmions start
to localize below some temperature. \textbf{}For higher temperatures
the number of skyrmions will be only determined by the deviation of
the filling factor from $\nu =1$. Therefore, around any additional
electron or hole (whose number is $N_{0}$) introduced in the completely
polarized state, a new spin texture will be formed. In this case the
number of free, or moving, skyrmions \textbf{($N_{fs}$)} exactly
coincides with $N_{0}$, which is related to the electronic density
of the 2DEG $n_{0}$ by$N_{0}=n_{0}\mathcal{S}\left|1-\nu \right|$. 

As it can be seen our model assume that as the temperature is lowered
the number of free skyrmions decreases. At first sight if the unbinding
of skyrmions from the disordered centers is a thermally activated
process, the \textbf{}number of free spin textures will be roughly
given by $N_{sf}=n_{0}\mathcal{S}\left|1-\nu \right|\exp \left(-U/T\right)$,
where $U$ is some average value of the attractive potential induced
by the disorder. However, this simple assumption implies that all
particles localize at the same temperature, in contradiction with
the different patterns observed in the Knight shift measurements for
different values of $\nu $. \textbf{}In order to be more realistic
in modeling the disorder, a certain degree of correlation between
them should be included. The main effect of this correlation is nothing
but a distribution of potentials with different depths in such a way
that as the temperature decreases the skyrmions start to localize
in the deepest wells and this process continues until the {}``last
skyrmion'' localizes in the shallowest well. If this distribution
is assumed to be Gaussian around some specific value $U_{o}$, the
number of potentials with depth $U$ can be written as \textbf{\begin{equation}
n(U)=\mathcal{N}exp\left(-(U-U_{0})^{2}/\Delta ^{2}\right),\label{eq:distribu}\end{equation}
} where the constant $\mathcal{N}$ is related to the total number
of donors localized in the barriers between the GaAs quantum wells.
As all the electrons in the 2DEG arise from the donors their number
should be equal, i.e.\textbf{,}\begin{eqnarray}
n_{0}\mathcal{S} & = & \mathcal{N}\int _{0}^{\infty }\; dU\, e^{-\left(U-U_{0}\right)^{2}/\Delta ^{2}}\nonumber \\
 &  & \nonumber \\
 & = & \frac{\sqrt{\pi }}{2}\mathcal{N}\Delta \left(1+erf\left(U_{0}/\Delta \right)\right),\label{eq:constante1}
\end{eqnarray}
 where $erf(x)$ is the error integral \cite{Gradshteyn}.

Now, using expressions (\ref{eq:difcooper}) and (\ref{eq:distribu})
the temperature dependence of the transversal relaxation time (\ref{eq:tesegunda})
can be generalized to be a mean value over the potential distribution
as\begin{equation}
T_{2}^{-1}\sim \frac{e^{2g/T}}{T^{3}}\int _{U_{min}}^{\infty }\; dU\, e^{-\left(U-U_{0}\right)^{2}/\Delta ^{2}}e^{-U/T}.\label{eq:t2intermed}\end{equation}
Here, the value of $U_{min}$ is related to the total number of skyrmions
in the 2DEG through \begin{eqnarray}
n_{0}\left|1-\nu \right|\mathcal{S} & = & \mathcal{N}\int _{U_{min}}^{\infty }\; dU\, e^{-\left(U-U_{0}\right)^{2}/\Delta ^{2}}\nonumber \\
 &  & \nonumber \\
 & = & \frac{\sqrt{\pi }}{2}\mathcal{N}\Delta \left[1+erf\left(\frac{U_{min}-U_{0}}{\Delta }\right)\right].\label{eq:constante2}
\end{eqnarray}
Notice that as the number of skyrmions increases with$\left|1-\nu \right|$,
the value of $U_{min}$ should decrease in order to accommodate all
skyrmions in the localization centers.

Finally, integrating Eq. (\ref{eq:t2intermed}) and using the expressions
(\ref{eq:constante1}) and (\ref{eq:constante2}), it is possible
to show that \begin{equation}
T_{2}^{-1}\sim \frac{e^{2g/T-(\Delta ^{2}-4TU_{0})/4T^{2}}}{T^{3}}\left[1-erf\left(\frac{\Delta }{2T}+\phi \right)\right]\label{eq:t2final}\end{equation}
 where $\phi $ is given by \[
\phi =erf^{-1}\left(1-\left|1-\nu \right|\left[1+erf\left(\frac{U_{0}}{\Delta }\right)\right]\right),\]
 and $erf^{-1}(x)$ is the inverse function of $erf(x)$. 

As it can be checked expression (\ref{eq:t2final}) is a non monotonic
function of the temperature. In fact in the high temperature limit,
where the localization effects are negligible, the NMR linewidth ($T_{2}^{-1}$)
goes to zero in agreement with the motional narrowed effect observed
in the experiments. If we look in the opposite direction (low temperatures)
where the slow motion of the skyrmions tends to increase the linewidth,
the exponentially small number of free skyrmions cancels out this
effect and the linewidth decreases again. This is precisely the frozen
limit found in the NMR measurements. Therefore the interplay between
skyrmion dynamics and localization induced by disorder in the system
reproduces some of the major features of the transversal relaxation
time near $\nu =1$. It is worth pointing out that depending on the
disorder strength the 2DEG, away from $\nu =1$, can evolve to different
phases\cite{skydisorder} which in principle can influence the linewidth
measurements. However, these physics was not included in our very
simple model.

\section{Skyrmion in the presence of disorder}

At this point nothing is left but to estimate the value of $U_{0}$
due to the impurities localized outside the quantum well. This can
be done computing the change in energy of an isolated skyrmion due
to the presence of a positive charged ion. In order to do that the
skyrmions will be described by an effective generalized nonlinear
sigma model \cite{skyrmions-first} in terms of an unit vector field
$\mathbf{m}(\mathbf{r})$ associated to the electronic spin orientation.
The Lagrangian density which describes these objects in the presence
of an external magnetic field $\mathbf{B}$ can be written as \begin{eqnarray}
\mathcal{L}_{0} & = & T(\mathbf{m})-V(\mathbf{m}),\label{eq:sondhi}
\end{eqnarray}

where

\begin{equation}
T(\mathbf{m})=\frac{{\hbar \rho }}{4}\, \mathbf{A}(\mathbf{m})\cdot \partial _{t}\, \mathbf{m},\label{eq:kinetic}\end{equation}

and\begin{eqnarray}
V(\mathbf{m}) & = & \frac{{1}}{2}\Big [\, \rho _{s}\, (\nabla \mathbf{m})^{2}+g^{*}\overline{\rho }\mu _{B}\, \mathbf{m}\cdot \mathbf{B}\nonumber \\
 &  & -\, \frac{{e^{2}}}{\epsilon }\, \int dr'^{2}\: \frac{{q(r)\, q(r')}}{\left|\mathbf{r}-\mathbf{r}'\right|}\, \Big ].\label{eq:popi}
\end{eqnarray}
In the kinetic term (\ref{eq:sondhi}) $\rho $ denotes the electronic
density and $\mathbf{A}[\mathbf{m}]$ corresponds to the vector potential
of a unit monopole, i.e., $\epsilon ^{ijk}\partial _{j}A^{k}=m_{i}$.
On the other hand in the potential energy density (\ref{eq:popi}),
$\rho _{s}$ is the spin stiffness, $g^{*}$ is the effective Landé
factor, $\overline{\rho }=1/(2\pi l^{2})$ is the uniform electronic
background density and $\epsilon $ is the dielectric constant of
the background semiconductor. The deviation of the physical density
from the uniform value $\overline{\rho }$ determines the skyrmion
density $q(r)$, which can be explicitly written as\begin{equation}
q(r)=\frac{{1}}{8\pi }\: \epsilon _{\mu \eta }\, \mathbf{m}(\mathbf{r})\, \cdot \, (\partial _{\mu }\mathbf{m}(\mathbf{r})\: \times \: \partial _{\eta }\mathbf{m}(\mathbf{r})),\label{eq:skydensity}\end{equation}
 and whose spatial integral is the topological charge.

\begin{figure}[t]
\begin{center}\includegraphics[  bb=4bp 24bp 596bp 532bp,
  clip,
  scale=0.33]{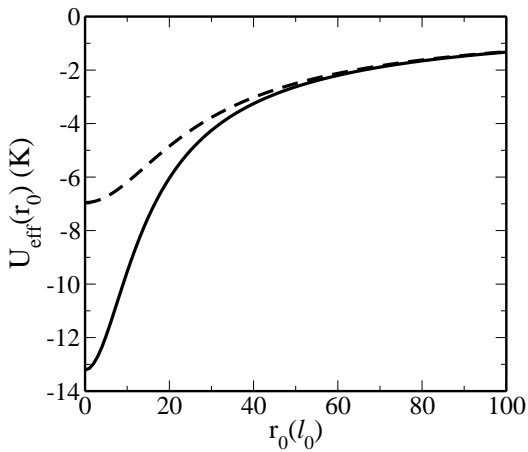}\end{center}

\caption{\label{figura2}Effective potential  in $K$ as a function of $\mathbf{r}_{0}$
( the skyrmion-donors distance in the the 2DEG). The continuous line
corresponds to $d=900\textrm{Å}$ and the dashed line to $d=1800\textrm{Å}$. }
\end{figure}

Instead of directly trying to solve (\ref{eq:sondhi})-(\ref{eq:popi}),
Sondhi \textit{et al.} \cite{skyrmions-first} considered the soliton
solution of the nonlinear $\sigma $ model obtained by Belavin and
Polyakov \cite{polyakov} with a fixed size $\lambda $, which is
set by the competition between the Coulomb and Zeeman interactions
included in (\ref{eq:popi}). Therefore, within this approach, the
static solution of (\ref{eq:sondhi}) can be written as \begin{equation}
m_{x(y)}^{0}=\frac{{4\lambda r\cos \theta (\sin \theta )}}{r^{2}+4\lambda ^{2}},\; \; \; \; m_{z}^{0}=\frac{{r^{2}-4\lambda ^{2}}}{r^{2}+4\lambda ^{2}},\label{eq:polyakov}\end{equation}
which describes a skyrmion of unit topological charge localized at
the origin. The skyrmion size is given by $\overline{\lambda }=0.558l_{0}(\widetilde{g}\left|\ln \widetilde{g}\right|)^{-1/3}$,
where $l_{0}$ stands for the magnetic length and $\widetilde{g}=g\mu _{B}B/\epsilon $.
In this approximation the skyrmion density is given by $q(r)={4\lambda ^{2}}/\pi (4\lambda ^{2}+r^{2})^{2}.$

As pointed out earlier, the disorder effect in the 2DEG is related
to the donors situated in a layer whose distance from the 2DEG is
$d$. In order to include those effects in the skyrmion dynamics,
we add an extra term to the the Lagrangian density (\ref{eq:sondhi}).
If we consider that the skyrmion density $q(\mathbf{r})$ is coupled
to the disorder potential via Coulomb interaction, our model for the
skyrmion-impurity system is given by the Lagrangian density \begin{equation}
\mathcal{L}=\mathcal{L}_{0}-\frac{{e^{2}}q(r)}{\epsilon \left|d^{2}+(\mathbf{r}-\mathbf{r}_{0})^{2}\right|}\label{eq:1impurity}\end{equation}
where $e$ is the electron charge and $\mathbf{r}_{0}$ denotes the
donor coordinate at the 2DEG plane. Once we are assuming a very simple
model to describe the system, we will not consider the fact that the
disorder potential can be screened by the 2DEG as pointed out by Efros
\emph{et al}\cite{efrosprb1}. 

>From expressions (\ref{eq:sondhi})-(\ref{eq:popi}) and (\ref{eq:1impurity}),
the energy functional for a single static skyrmion can be written
as \begin{eqnarray}
E & = & \int \Big [\, V[\mathbf{m}]+\frac{{e^{2}}q(\mathbf{r})}{\epsilon \left|d^{2}+(\mathbf{r}-\mathbf{r}_{0})^{2}\right|}\Big ]dr^{2},\label{eq:energiaefetiva}
\end{eqnarray}
where $V(\mathbf{m})$ is the interacting potential related to the
Lagrangian density (\ref{eq:sondhi})-(\ref{eq:popi}) and the integral
of the second term is performed over the whole plane containing the
2DEG. If we assume that the presence of the disorder does not modify
the skyrmion form and size, the energy of the system will be given
by Equation (\ref{eq:energiaefetiva}) with $\mathbf{m}(\mathbf{r})=\mathbf{m}^{0}(\mathbf{r})$.
Therefore, the change in energy of a single skyrmion due to the disorder
potential is given by $U_{eff}(\mathbf{r}_{0})=E-E_{0}$, where $E_{0}=\int V(\mathbf{m}^{0}(\mathbf{r},\overline{\lambda }))dr^{2}$.

The specific form of $U_{eff}$ as a function of $r_{0}$ (the skyrmion-donor
distance in the plane of the 2DEG) is illustrated in Fig. \ref{figura2}
for $\lambda =1.2l_{0}$ (the skyrmion size at $B=7T$, see appendix).
As it can be seen, the minimum in energy corresponds to the case where
the impurity position in the 2DEG plane exactly coincides with the
skyrmion center. Therefore, the parameter $U_{0}$ in Eq. (\ref{eq:t2final})
can be estimated as the value of $U_{eff}$ at $r_{0}=0$. If we use
a typical value of the distance from the donors to the 2DEG as $d\sim 1.800\, \textrm{Å}$
\cite{skyloc1ter}, we conclude that $U_{0}\sim 7\, K$. It is interesting
to notice that the value of the estimated $U_{0}$ is of the same
order of the disordered localizing potential reported in Ref. \onlinecite{nederveen}
for the case in which $\nu \approx 1$.

\section{Results and Discussion}

\begin{figure}[b]
\begin{center}\includegraphics[  bb=39bp 35bp 643bp 523bp,
  clip,
  scale=0.32]{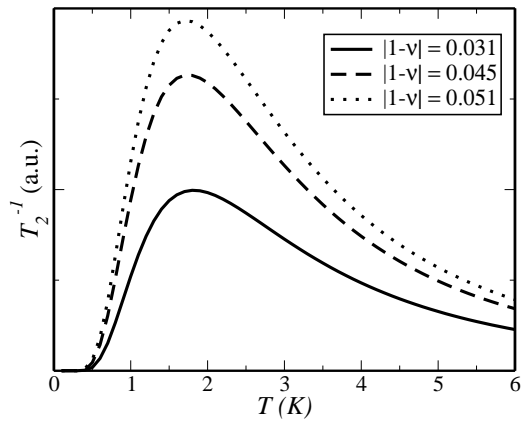}\end{center}

\caption{\label{figura} $T_{2}^{-1}$ (FWHM) as a function of temperature
for $U_{0}=7\, K$, $\Delta =2\, K$, $g=2\, K$ and different filling
factor values.}
\end{figure}

At this point we can turn or attention to the linewidth behavior as
a function of the temperature. Fig. \ref{figura} shows a plot of
equation (\ref{eq:t2final}) as a function of temperature for different
values of $\left|1-\nu \right|$ with $U_{0}=7\, K$, $\Delta =2\, K$
and $g=2\, K$ (the Zeeman gap at $B=7\, T$). One can see that, $T_{2}^{-1}$
has the non-monotonic behavior observed in the OPNMR experiments \cite{skyloc1,skyloc1ter}
going from the \emph{motional-narrowed} regime at relative high temperatures
to the \emph{frozen limit} as the temperature decreases. The agreement
between the theory and the experimental data reported in Ref. \onlinecite{skyloc1}
suggests that the relevant gapless mode that induce short nuclear
relaxation times $T_{1}$ near $\nu \approx 1$ is precisely the translational
mode \cite{skyavf,skyr-cooper}. On the other hand, it can be seen
that as $\mid 1-\nu \mid $ increases the maximum value of $1/T_{2}$
also increase. On physical grounds this can be expected, because as
the number of skyrmions grows, the possibility of finding \emph{free}
skyrmions which effectively induce nuclei dephasing, increases. Although
for the $\nu \approx 1$ measurements \cite{skyloc1} this kind of
behavior is almost unclear, in the $\nu \approx 1/3$ situation it
can be neatedly observed \cite{skyloc1ter}. 

For $U_{0}$ equal to the potential depth estimated from the isolated
impurity case, the peak position -indicating the change in the dynamical
regime- is around $1.7\, K$ which is very close to the reported experimental
data\cite{skyloc1}. It is possible to show that if $U_{0}$ changes
from $4\, K$ to $8\, K$, the peak position varies from $1\, K$
to $2.2\, K$ remaining within the reported results\cite{skyloc1}.
As it can be seen in Fig. \ref{figura}, the peak position is almost
constant as the filling factor changes, which is in contradiction
with the experimental results for $\nu \approx 1$\cite{skyloc1}
but in agreement with the $\nu \approx 1/3$ case\cite{skyloc1ter}.
This discrepancy could be possibly associated with the non interacting
skyrmions approach we have employed, that is more appropriate for
the small skyrmion near $\nu =1/3$. This conjecture can be tested
by applying an in-plane magnetic field, which does not change the
filling factor but increase the Zeeman energy making the skyrmion
smaller. 

It is worth pointing out that our model for the linewidth of the NMR
signal is in agreement with the scenario suggested for the observed
tilted plateau in the in the Knight shift measurements for the 2DEG
when $\nu \thickapprox 1$. In this case, the existence of the tilted
plateau is also related to the localization of the quasiparticles
introduced in the 2DEG as the filling factor deviates from $\nu =1$\cite{skyloc1}. 

Summarizing, we have developed a very simple theory which describes
the evolution of the NMR line profile of the $^{71}$Ga nuclei coupled
to the 2DEG (near $\nu =1$) as a function of the whole available
temperature range which seems to be in fairly good agreement with
the experimental data.

\section{Acknowledgments}

AVF would like to thank A.R. Pereira and M.O. Goerbig for helpful
discussions. The authors kindly acknowledge the financial support
from Fundação de Amparo à Pesquisa do Estado de São Paulo (FAPESP)
and AOC the partial support from Conselho Nacional de Desenvolvimento
Científico e Tecnológico (CNPq).

\section{Appendix \label{ap:energia}}

In the table below, we show the energy scales for the quantum Hall
system at $\nu =1$ in Kelvin. In all expressions, the magnetic field
is measure in Tesla. For GaAs quantum wells $\epsilon \approx 13$.

\begin{center}\begin{tabular}{|lrc|rr|}
\hline 
Energy scales &
\hspace{1.5cm}&
&
 \hspace{4.0cm}&
 (K) \\
\hline
 $\epsilon _{C}$&
&
$e^{2}/\epsilon l_{0}$&
&
$50.40\sqrt{B}$\\
$g$&
&
$g^{*}\mu _{B}B$&
&
$0.33B$\\
 $\rho _{s}$&
&
$\epsilon _{C}/(16\sqrt{2\pi })$&
&
$1.25\sqrt{B}$\\
\hline
\end{tabular}\end{center}

Therefore the skyrmion size $\lambda =0.558l_{0}(\widetilde{g}\left|\ln \widetilde{g}\right|)^{-1/3}$
calculated as described in the text can be written as $\lambda =3.08\left(\sqrt{B}\left|\ln \left(5.95\times 10^{-3}\sqrt{B}\right)\right|\right)^{-1/3}$,
where $l_{0}=\sqrt{\hbar c/eB}=256/\sqrt{B}$$\textrm{Å}$.

\bibliographystyle{apsrev}
\bibliography{t2}

\begin{thebibliography}{21}
\expandafter\ifx\csname natexlab\endcsname\relax\def\natexlab#1{#1}\fi
\expandafter\ifx\csname bibnamefont\endcsname\relax
  \def\bibnamefont#1{#1}\fi
\expandafter\ifx\csname bibfnamefont\endcsname\relax
  \def\bibfnamefont#1{#1}\fi
\expandafter\ifx\csname citenamefont\endcsname\relax
  \def\citenamefont#1{#1}\fi
\expandafter\ifx\csname url\endcsname\relax
  \def\url#1{\texttt{#1}}\fi
\expandafter\ifx\csname urlprefix\endcsname\relax\def\urlprefix{URL }\fi
\providecommand{\bibinfo}[2]{#2}
\providecommand{\eprint}[2][]{\url{#2}}

\bibitem[{\citenamefont{Girvin}(1999)}]{girvin}
\bibinfo{author}{\bibfnamefont{S.}~\bibnamefont{Girvin}},
  \bibinfo{journal}{cond-mat/9907002}  (\bibinfo{year}{1999}).

\bibitem[{\citenamefont{{S. L. Sondhi} et~al.}(1993)\citenamefont{{S. L.
  Sondhi}, {A. Karlhede}, {S. A. Kivelson}, and {E.H.
  Rezayi}}}]{skyrmions-first}
\bibinfo{author}{\bibnamefont{{S. L. Sondhi}}},
  \bibinfo{author}{\bibnamefont{{A. Karlhede}}},
  \bibinfo{author}{\bibnamefont{{S. A. Kivelson}}}, \bibnamefont{and}
  \bibinfo{author}{\bibnamefont{{E.H. Rezayi}}}, \bibinfo{journal}{Phys. Rev.
  B} \textbf{\bibinfo{volume}{47}}, \bibinfo{pages}{16419}
  (\bibinfo{year}{1993}).

\bibitem[{\citenamefont{{K. Lejnell} et~al.}(1999)\citenamefont{{K. Lejnell},
  {A. Karhede}, and {S.L. Sondhi}}}]{sky-size}
\bibinfo{author}{\bibnamefont{{K. Lejnell}}}, \bibinfo{author}{\bibnamefont{{A.
  Karhede}}}, \bibnamefont{and} \bibinfo{author}{\bibnamefont{{S.L. Sondhi}}},
  \bibinfo{journal}{Phys. Rev. B} \textbf{\bibinfo{volume}{59}},
  \bibinfo{pages}{10 183} (\bibinfo{year}{1999}).

\bibitem[{\citenamefont{{S.E. barret} et~al.}(1995)\citenamefont{{S.E. barret},
  {G. Dabbagh}, {L.N. Pfeiffer}, {K.W. West}, and {R. Tycko}}}]{skyevidence}
\bibinfo{author}{\bibnamefont{{S.E. barret}}},
  \bibinfo{author}{\bibnamefont{{G. Dabbagh}}},
  \bibinfo{author}{\bibnamefont{{L.N. Pfeiffer}}},
  \bibinfo{author}{\bibnamefont{{K.W. West}}}, \bibnamefont{and}
  \bibinfo{author}{\bibnamefont{{R. Tycko}}}, \bibinfo{journal}{Phys. Rev.
  Lett.} \textbf{\bibinfo{volume}{74}}, \bibinfo{pages}{5112}
  (\bibinfo{year}{1995}).

\bibitem[{\citenamefont{{V. Bayot} et~al.}(1996)\citenamefont{{V. Bayot}, {E.
  Grivei}, {S. Melinte}, {M.B. Santos}, and {M. Shayegan}}}]{skyheat}
\bibinfo{author}{\bibnamefont{{V. Bayot}}}, \bibinfo{author}{\bibnamefont{{E.
  Grivei}}}, \bibinfo{author}{\bibnamefont{{S. Melinte}}},
  \bibinfo{author}{\bibnamefont{{M.B. Santos}}}, \bibnamefont{and}
  \bibinfo{author}{\bibnamefont{{M. Shayegan}}}, \bibinfo{journal}{Phys. Rev.
  Lett.} \textbf{\bibinfo{volume}{76}}, \bibinfo{pages}{4584}
  (\bibinfo{year}{1996}).

\bibitem[{\citenamefont{{E.H. Aifer} et~al.}(1996)\citenamefont{{E.H. Aifer},
  {B.B. Goldberg}, and {D.A. Broido}}}]{skymagabs1}
\bibinfo{author}{\bibnamefont{{E.H. Aifer}}},
  \bibinfo{author}{\bibnamefont{{B.B. Goldberg}}}, \bibnamefont{and}
  \bibinfo{author}{\bibnamefont{{D.A. Broido}}}, \bibinfo{journal}{Phys. Rev.
  Lett.} \textbf{\bibinfo{volume}{76}}, \bibinfo{pages}{680}
  (\bibinfo{year}{1996}).

\bibitem[{\citenamefont{{A. Schmeller} et~al.}(1995)\citenamefont{{A.
  Schmeller}, {J.P. Eisenstein}, {L.N. Pfeiffer}, and {K.W. West}}}]{skytrans1}
\bibinfo{author}{\bibnamefont{{A. Schmeller}}},
  \bibinfo{author}{\bibnamefont{{J.P. Eisenstein}}},
  \bibinfo{author}{\bibnamefont{{L.N. Pfeiffer}}}, \bibnamefont{and}
  \bibinfo{author}{\bibnamefont{{K.W. West}}}, \bibinfo{journal}{Phys. Rev.
  Lett.} \textbf{\bibinfo{volume}{75}}, \bibinfo{pages}{4290}
  (\bibinfo{year}{1995}).

\bibitem[{\citenamefont{{P. Khandelwal} et~al.}(2001)\citenamefont{{P.
  Khandelwal}, {A.E. Dementyev}, {N.N. Kuzma}, {S.E. Barret}, {L.N. Pfeiffer},
  and {K.W. West}}}]{skyloc1}
\bibinfo{author}{\bibnamefont{{P. Khandelwal}}},
  \bibinfo{author}{\bibnamefont{{A.E. Dementyev}}},
  \bibinfo{author}{\bibnamefont{{N.N. Kuzma}}},
  \bibinfo{author}{\bibnamefont{{S.E. Barret}}},
  \bibinfo{author}{\bibnamefont{{L.N. Pfeiffer}}}, \bibnamefont{and}
  \bibinfo{author}{\bibnamefont{{K.W. West}}}, \bibinfo{journal}{Phys. Rev.
  Lett.} \textbf{\bibinfo{volume}{86}}, \bibinfo{pages}{5353}
  (\bibinfo{year}{2001}).

\bibitem[{\citenamefont{{C.P. Slichter}}(1990)}]{slichter}
\bibinfo{author}{\bibnamefont{{C.P. Slichter}}},
  \emph{\bibinfo{title}{Principles of {M}agnetic {R}esonance}}
  (\bibinfo{publisher}{Springer-Verlag}, \bibinfo{year}{1990}).

\bibitem[{\citenamefont{{N.N. Kuzma} et~al.}(1998)\citenamefont{{N.N. Kuzma},
  {P. Khandelwal}, {S.E. Barret}, {L.N. Pfeiffer}, and {K.W.
  West}}}]{skyloc1ter}
\bibinfo{author}{\bibnamefont{{N.N. Kuzma}}}, \bibinfo{author}{\bibnamefont{{P.
  Khandelwal}}}, \bibinfo{author}{\bibnamefont{{S.E. Barret}}},
  \bibinfo{author}{\bibnamefont{{L.N. Pfeiffer}}}, \bibnamefont{and}
  \bibinfo{author}{\bibnamefont{{K.W. West}}}, \bibinfo{journal}{Science}
  \textbf{\bibinfo{volume}{281}}, \bibinfo{pages}{686} (\bibinfo{year}{1998}).

\bibitem[{\citenamefont{{J. Sinova}
  et~al.}(2000{\natexlab{a}})\citenamefont{{J. Sinova}, {S.M. Girvin}, {T.
  Jungwirth}, and {K. Moon}}}]{nmrlineshape}
\bibinfo{author}{\bibnamefont{{J. Sinova}}},
  \bibinfo{author}{\bibnamefont{{S.M. Girvin}}},
  \bibinfo{author}{\bibnamefont{{T. Jungwirth}}}, \bibnamefont{and}
  \bibinfo{author}{\bibnamefont{{K. Moon}}}, \bibinfo{journal}{Phys. Rev. B}
  \textbf{\bibinfo{volume}{61}}, \bibinfo{pages}{2749}
  (\bibinfo{year}{2000}{\natexlab{a}}).

\bibitem[{\citenamefont{{L. Brey} et~al.}(1995)\citenamefont{{L. Brey}, {H.A.
  Fertig}, {R. C\^ot\'e}, and {A.H. MacDonald}}}]{skycrystal}
\bibinfo{author}{\bibnamefont{{L. Brey}}}, \bibinfo{author}{\bibnamefont{{H.A.
  Fertig}}}, \bibinfo{author}{\bibnamefont{{R. C\^ot\'e}}}, \bibnamefont{and}
  \bibinfo{author}{\bibnamefont{{A.H. MacDonald}}}, \bibinfo{journal}{Phys.
  Rev. Lett.} \textbf{\bibinfo{volume}{75}}, \bibinfo{pages}{2562}
  (\bibinfo{year}{1995}).

\bibitem[{\citenamefont{{A. Villares Ferrer} and {A.O.
  Caldeira}}(2000)}]{skyavf}
\bibinfo{author}{\bibnamefont{{A. Villares Ferrer}}} \bibnamefont{and}
  \bibinfo{author}{\bibnamefont{{A.O. Caldeira}}}, \bibinfo{journal}{Phys. Rev.
  B} \textbf{\bibinfo{volume}{61}}, \bibinfo{pages}{2755}
  (\bibinfo{year}{2000}).

\bibitem[{\citenamefont{{A.G. Green} and {N.R. Cooper}}(2002)}]{skyr-cooper}
\bibinfo{author}{\bibnamefont{{A.G. Green}}} \bibnamefont{and}
  \bibinfo{author}{\bibnamefont{{N.R. Cooper}}}, \bibinfo{journal}{Phys. Rev.
  B} \textbf{\bibinfo{volume}{65}}, \bibinfo{pages}{10 183}
  (\bibinfo{year}{2002}).

\bibitem[{\citenamefont{{R. Rajaraman}}(1989)}]{rajaraman}
\bibinfo{author}{\bibnamefont{{R. Rajaraman}}}, \emph{\bibinfo{title}{Solitons
  and {I}nstantons}} (\bibinfo{publisher}{North-Holland, Amterdam},
  \bibinfo{year}{1989}).

\bibitem[{\citenamefont{{R.P. Feynman} and {F.L.
  Vernon}}(1963)}]{feynmanvernon}
\bibinfo{author}{\bibnamefont{{R.P. Feynman}}} \bibnamefont{and}
  \bibinfo{author}{\bibnamefont{{F.L. Vernon}}}, \bibinfo{journal}{Ann. Phys.
  (N.Y.)} \textbf{\bibinfo{volume}{24}}, \bibinfo{pages}{118}
  (\bibinfo{year}{1963}).

\bibitem[{\citenamefont{{A.J. Nederveen} and {Y.V. Nazarov}}(1999)}]{nederveen}
\bibinfo{author}{\bibnamefont{{A.J. Nederveen}}} \bibnamefont{and}
  \bibinfo{author}{\bibnamefont{{Y.V. Nazarov}}}, \bibinfo{journal}{Phys. Rev.
  Lett.} \textbf{\bibinfo{volume}{82}}, \bibinfo{pages}{406}
  (\bibinfo{year}{1999}).

\bibitem[{\citenamefont{Gradshteyn and Ryzhik}(1980)}]{Gradshteyn}
\bibinfo{author}{\bibfnamefont{I.~S.} \bibnamefont{Gradshteyn}}
  \bibnamefont{and} \bibinfo{author}{\bibfnamefont{I.~M.}
  \bibnamefont{Ryzhik}}, \emph{\bibinfo{title}{Table of Integral Series and
  Products}} (\bibinfo{publisher}{Academic Press Inc.}, \bibinfo{year}{1980}),
  \bibinfo{edition}{4th} ed.

\bibitem[{\citenamefont{{J. Sinova}
  et~al.}(2000{\natexlab{b}})\citenamefont{{J. Sinova}, {A.H. MacDonald}, and
  {S.M. Girvin}}}]{skydisorder}
\bibinfo{author}{\bibnamefont{{J. Sinova}}},
  \bibinfo{author}{\bibnamefont{{A.H. MacDonald}}}, \bibnamefont{and}
  \bibinfo{author}{\bibnamefont{{S.M. Girvin}}}, \bibinfo{journal}{Phys. Rev.
  B} \textbf{\bibinfo{volume}{62}}, \bibinfo{pages}{13579}
  (\bibinfo{year}{2000}{\natexlab{b}}).

\bibitem[{\citenamefont{{A.A. Belavin} and {A.M. Polyakov}}(1975)}]{polyakov}
\bibinfo{author}{\bibnamefont{{A.A. Belavin}}} \bibnamefont{and}
  \bibinfo{author}{\bibnamefont{{A.M. Polyakov}}}, \bibinfo{journal}{Pis'ma Zh.
  \'Eksp. Teor. Fiz.} \textbf{\bibinfo{volume}{22}}, \bibinfo{pages}{503}
  (\bibinfo{year}{1975}).

\bibitem[{\citenamefont{{A.L. Efros} et~al.}(1993)\citenamefont{{A.L. Efros},
  {F.G. Pikus}, and {V.G. Burnett}}}]{efrosprb1}
\bibinfo{author}{\bibnamefont{{A.L. Efros}}},
  \bibinfo{author}{\bibnamefont{{F.G. Pikus}}}, \bibnamefont{and}
  \bibinfo{author}{\bibnamefont{{V.G. Burnett}}}, \bibinfo{journal}{Phys. Rev.
  B} \textbf{\bibinfo{volume}{47}}, \bibinfo{pages}{2233}
  (\bibinfo{year}{1993}).

\end{thebibliography}

\end{document}